# Precise nanosizing with high dynamic range holography

Unai Ortiz-Orruño[1], Ala Jo[2,3], Hakho Lee[2,3], Niek F. van Hulst[1,4] and Matz Liebel[1,*]

[1] ICFO -Institut de Ciencies Fotoniques, The Barcelona Institute of Science and Technology, 08860 Castelldefels, Barcelona, Spain

[2] Center for Systems Biology, Massachusetts General Hospital, Boston, Massachusetts 02114, USA

[3] Department of Radiology, Massachusetts General Hospital, Boston, Massachusetts 02114, USA

[4] ICREA -Institució Catalana de Recerca i Estudis Avançats, 08010 Barcelona, Spain

\* Matz.Liebel@ICFO.eu

Optical sensing is one of the key-enablers of modern diagnostics. Especially label-free imaging modalities hold great promise as they eliminate labelling procedures prior to analysis. However, scattering signals of nanometric particles scale with their volume-square. This unfavourable scaling makes it extremely difficult to quantitatively characterise intrinsically heterogeneous clinical samples, such as extracellular vesicles, as their signal variation easily exceeds the dynamic range of currently available cameras. Here, we introduce off-axis k-space holography that circumvents this limitation. By imaging the back-focal-plane of our microscope we project the scattering signal of all particles onto all camera pixels thus dramatically boosting the achievable dynamic range to up-to 110 dB. We validate our platform by detecting, and quantitatively sizing, metallic and dielectric particles over a 200x200 µm field-of-view and demonstrate that independently performed signal calibrations allow correctly sizing particles made from different materials. Finally, we present quantitative size-distributions of extracellular vesicle samples.

Holography, Extracellular Vesicles, Label-Free, Interferometric Microscopy, Fourier Imaging

**Motivation,**

Diagnostics is key to detecting, quantifying and, ultimately, curing diseases. A very important aspect of modern diagnostics is to provide rapid identification of a medical condition at the highest possible sensitivity. Ideally, a novel platform yields quantitative results that allow taking necessary measures to prevent the spread of any infectious disease or to commence a suitable treatment before non-reversible conditions develop. Numerous, promising approaches to providing rapid diagnostics are based on label-free detection which eliminates the need for sample labelling prior to analysis, thus dramatically reducing the overall time needed.

An extremely promising approach to robust particle quantification is digital holography which is intrinsically label-free. Furthermore, it allows replacing large fractions of complex imaging systems by computational operations. For macroscopic samples, such as tissue samples or cell populations, lensless imaging can provide diffraction limited resolution, over millimetre-sized field-of-views, thus paving the way towards next-generation point-of-care devices[1,2]. On the nanoscale, lens-based inline holography under various acronyms such as: IRM[3], IRIS[4], COBRI[5], iSCAT[6], iSCAMs[7] or stroboSCAT[7], to name a few, enables high-speed observations, in the hundreds of kilohertz range, and highly-sensitive label-free particle size estimation[7–10]. However, a big problem is the dramatic scaling of the scattering signal with particle size. This dependence is especially problematic for heterogeneous biological samples such as extracellular vesicles (EVs)[11–13], or exosomes, which span diameter-ranges of, approximately, 20-250 nm[14,15].

**Scattering signal considerations,**

Even though our daily intuition suggests that it should be trivial to, simultaneously, observe objects which differ by approximately twenty in size on the nanoscale this intuition fails dramatically. In the sub-wavelength regime, a particle's scattering amplitude scales with its volume. As we necessarily detect electric field intensities the resulting signal, $I_{sca}$, scales with the volume, $V$, square:

$$I_{sca} = |E_{sca}|^2 \propto V^2 \propto d^6 \qquad (1)$$

In other words, the difference between a, for example, 10 nm and a 200 nm particle is $20^6 \sim 10^8$, if we assume Rayleigh scattering. Darkfield-based quantification of said particle range is extremely challenging and widefield imaging is outright impossible as currently available cameras do not cover the required dynamic range. Holographic approaches, where the scattered field, $E_{sca}$ is interfered with a reference wave, $E_{ref}$, mitigate some of the problems as they allow signal amplification in a typical heterodyne-fashion thus eliminating the need for unrealistically low detector dark-counts. The holographic signal can be described as:

$$I_{hol} = |E_{ref}|^2 + |E_{sca}|^2 + 2E_{sca}E_{ref}\cos[\Delta\varphi] \qquad (2)$$

, where $|E_{sca}|^2$ is the scattering intensity described above, $|E_{ref}|^2$ the reference intensity and $2E_{sca}E_{ref}\cos[\Delta\varphi]$ the interference term of the two electric fields which depends on their phase difference. In the limit of strong particle scattering, $E_{sca} \gg E_{ref}$, we essentially recover darkfield behaviour and Equation 2 simplifies to Equation 1. Figure 1a compares the holographically obtained intensity, for two different reference-wave amplitudes, to the volume-square limit discussed above. Even though the total signal shows reference-wave dependent differences, the absolute signal variation remains, essentially, unaltered close to the factor of approximately $10^8$ for darkfield scattering. Furthermore, the simultaneous detection of both the amplitude square and the interference terms results in particle-size ambiguities as different diameters exhibit the same signal (Figure 1a, inset).

**Ideal sensing platform,**

An ideal optical sensing platform for heterogeneous samples such as EV populations hence: i) requires an extreme dynamic range, ii) isolates the interference term from the total signal to avoid ambiguities and, iii) allows large-field-of-view observations to build the necessary statistics that enable quantitative observations.

**Experimental implementation,**

Here, we implement such a sizing platform in the form of an off-axis k-space holographic microscope (k-scope). The k-scope relies on off-axis interference in momentum space[16,17], that is, it interferes a reference wave with an image of the back-focal-plane (BFP) of the microscope objective. As such, it yields holographic information that, ultimately, allows recovering a real-space image of the sample. This configuration has the advantage that it projects the scattering signal of all particles onto all camera pixels to increase the total dynamic range up to what would be achievable by using the entire detector as a single pixel. Contrary to a point-detector, such as a photodiode, our scheme employs off-axis holography[18,19] and allows recording large-field-of-view images, equivalent to those that would be recorded in a common real-space imaging system albeit a dramatically increased dynamic range where the improvement is directly proportional to the total number of camera pixels. A 9-megapixel CMOS camera, with an intrinsic dynamic range of 40 dB, can thus be boosted to 110 dB.

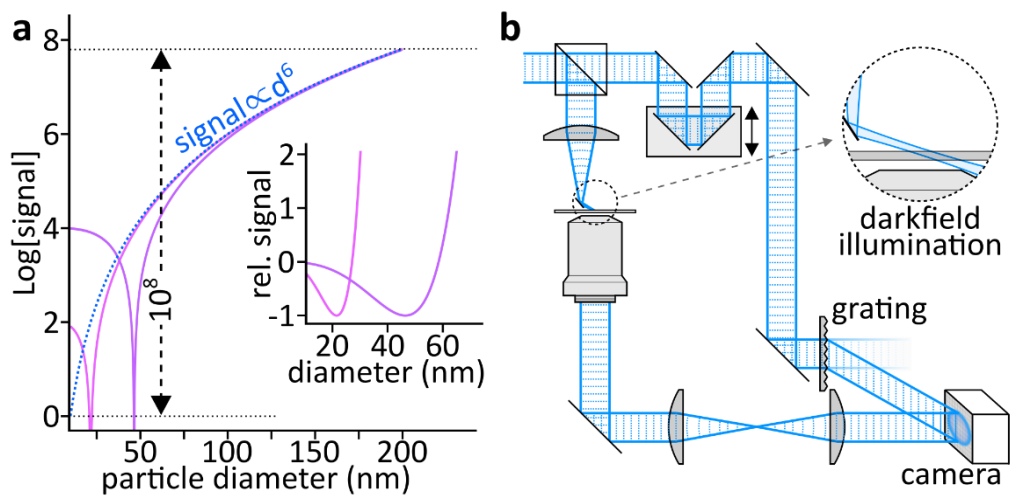

**Figure 1, Signal scaling at the nanoscale and experimental sensing.** a) LogLinear representation of the particle diameter dependent holographic signal intensities for different reference amplitudes (solid) assuming inline-detection and an $E_{ref}$ to $E_{sca}$ phase-shift of π. The dotted curve describes the limit of $E_{ref} \to 0$ which is equivalent to darkfield detection. The inset shows the normalised signal intensities around $E_{sca} \approx E_{ref}$ as $I_{hol}/|E_{ref}|^2 - 1$, on a linear scale, to highlight the signal ambiguity problem. b) Schematic of our off-axis holographic k-scope which employs a grating to match the phase-front and a delay-line to match the phase delay. The inset shows a magnification of the darkfield-mask free collection configuration.

Figure 1b shows a schematic of our k-scope where a beamsplitter generates both illumination and reference fields from a diode laser centred at 520 nm (*LDM-520-100-C, Lasertack*). The former illuminates the sample under an oblique illumination configuration (Figure 1b, inset). A microscope objective (*Mitutoyo MY100X-806*, NA=0.7) collects the sample scattering and a relay imaging system images the objective's back-focal-plane (BFP) onto a camera (*acA2040-90um, Basler AG*) where it interferes with the reference wave in an off-axis configuration. We, furthermore, use a mechanical delay-line and a grating, for phase-front matching, to ensure maximum interference contrast over the entire detector[a].

**Interferogram processing,**

Figure 2a shows an as-detected camera image obtained of a sample containing 20 nm diameter Au nanoparticles (NPs) that we illuminate at 520 nm under an oblique illumination angle of 60° with respect to the sample normal. The image exhibits a circular region, which is the relay-image of the BFP, and the superimposed reference wave which shows low frequency modulations due to the low-quality diffractive beamsplitter employed. A closer look additionally reveals high-frequency modulation which are a direct result of the off-axis interference of the two terms[19,20]. In other words, the off-axis configuration directly allows separating the interference term from the amplitude square contributions (Equation 2). The real-space image of the 20 nm Au NP-sample is directly obtained as the amplitude of the fast Fourier transformation of the recording (Figure 2a), where the image-information of interest is well-separated from the DC term located at the corners of our k-space representation. Most importantly, we are able to detect many individual 20 nm particles with simultaneously high signal-to-noise and signal-to-background ratios. Additionally, our oblique illumination scheme decouples the excitation light from the collection objective to allow very large field-of-views, of approximately 200x200 μm in this example, which are difficult to achieve in through-objective configurations due to parasitic back-reflections that contaminate the signal of interest. Moreover, it increases the dynamic range as more pixels collect scattering signal.

**Power dependence and reference amplitude,**

Systematic power-dependence measurements of both the illumination as well as the reference fields confirm the expected linear dependence of the interferometric signal (Equation 2) with respect to both the illumination as well as the reference beam-amplitudes (Figure 2b). Apart from verifying that the image is purely composed of interferometric contributions the scaling has important implications for selecting ideal illumination/reference powers. Both the signal, as well as shot noise, scale linearly with the reference-amplitude and the signal-to-noise-ratio should thus be independent of its amplitude. However, sample dependent factors might slightly impact this notion[21] and, most importantly, it is necessary to operate above the dark-noise of the camera which dictates the minimum amplitude of the reference wave. This minimum-reference approach conveniently reserves the highest possible camera range for detecting photons originating from the sample. The illumination power is then simply adjusted to ensure a sufficiently high signal-to-noise-ratio, without saturating the camera or destroying the sample.

---

[a] Adjustment of time delay and grating is only necessary for some, non-stabilised, diode lasers which might exhibit surprisingly short temporal coherence lengths (>100 μm). Diode pumped solid state lasers, however, do not require adjustments, as long as the path length difference is kept within a few tens of centimetres, which considerably simplifies the experimental setup.

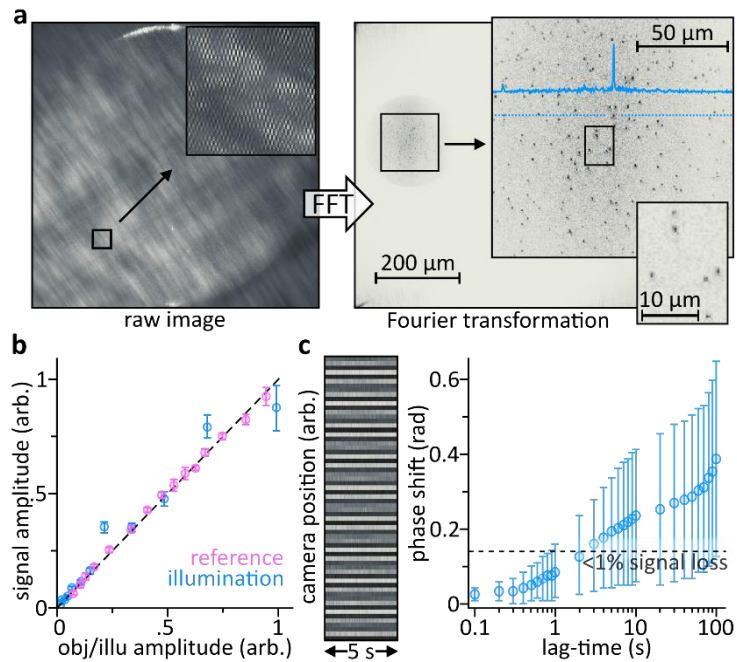

**Figure 2, Experimental implementation and data processing.** a) A typical back-focal-plane interferogram as recorded on the camera (left), for 20 nm diameter gold nanoparticles using an NA=0.7 air-objective. Right: The respective Fourier transformation reveals the image of interest over a large field-of-view; the amplitude range of the images with 200 µm and 50 µm scalebars has been restricted to approximately 20% of the maximum to visualise the particles. b) Signal amplitude dependence on the illumination (purple) and reference (blue) field-amplitudes. c) Interference fringe stability over 5s (left) and mean phase-shift for different lag-times between frames computed based on a 30 min experiment performed at 10 frames-per-second at 15 ms integration time (right). All error bars correspond to one standard deviation.

**Interferometric stability,**

One additional consideration is the interferometric stability where inline holography is often championed as being "more-stable" as the reference and scattering fields are intrinsically phase-locked. Based on our experience, however, off-axis holography only shows minor phase instabilities, that do not impact the signal quality, as long as sensible noise-reduction measures, such as floating the optical table, employing low beam heights and protecting the beam-path from air currents, are taken. To quantify phase instabilities under extreme conditions we separate the reference and illumination fields and independently propagated them over 3 m of path before recombining them in the typical off-axis configuration. Figure 2c shows a phase-drift measurement with no statistically significant deviations from zero phase-shift for lag-times of <400 ms and roughly 0.1 rad of drift, being equivalent to 8 nm, within one second. Such a phase drift manifests itself in a minor noise contribution on the order of 1% or, in other words, it is necessary to achieve unlikely single-particle signal-to-noise ratios exceeding 100 before the 0.1 rad phase-drift starts impacting the signal. More importantly, off-axis holography measures the signals' phase and computational phase-correction, prior to image averaging, is easily possible and eliminates the problem completely, as verified by performing signal-averaging experiments over observation times exceeding several tens of minutes.

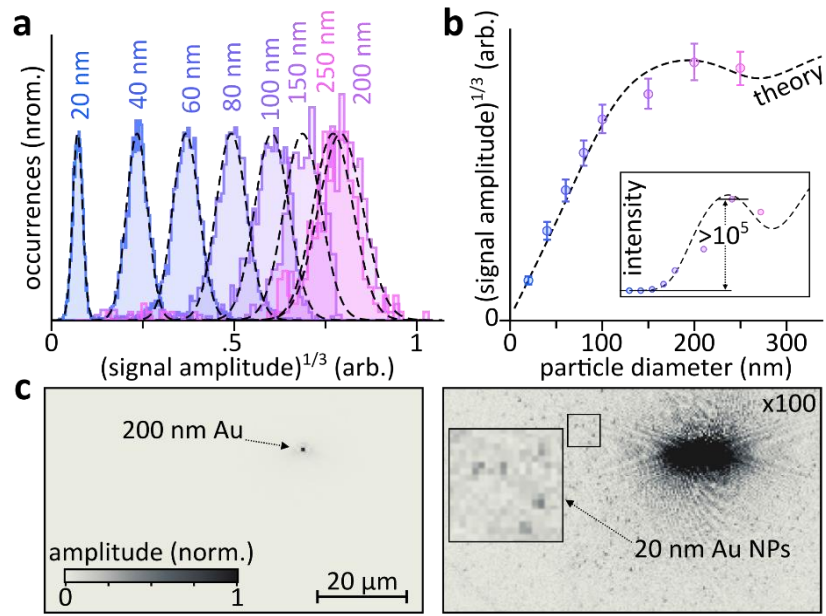

**Figure 3, Quantitative sizing of gold nanoparticles.** a) Signal distribution histograms measured for Au NPs of different sizes, 20-250 nm (coloured) alongside Gaussian fits (dashed). The histograms contain: $N_{20nm}$=2563, $N_{40nm}$=3709, $N_{60nm}$=3354, $N_{80nm}$=1758, $N_{100nm}$=676, $N_{150nm}$=329, $N_{200nm}$=288, $N_{250nm}$=243, b) Comparison of the experimentally obtained particle size dependent (signal amplitudes)$^{1/3}$ with the theoretically expected values assuming scattering collection angles of 75-165°, 520 nm illumination and gold particles surrounded by air with a constant refractive index of n=1. The error bars correspond to one standard deviation. Inset: Same data and simulation but using an intensity scale corresponding to the signal-levels detected on the camera. c) Amplitude image recorded for a sample containing 200 nm and 20 nm Au NPs normalised to the maximum (left) and to 1/100 of it (right). The faint line in the bottom left corner is due to steps in the mica surface.

**Gold nanoparticle detection,**

Following this thorough technical characterisation of the sensing platform, we now turn our attention to quantitative particle size-measurements. As an initial test system, we employ Au NPs of different nominal diameters ranging from 20 nm to 250 nm (*BBI Solutions*) that we immobilise on freshly cleaved, PLL-g-PEG -functionalised, mica (Methods). To demonstrate the extreme dynamic-range of the k-scope we perform all measurements under identical experimental conditions with the same illumination/reference intensities and camera-integration times. Figure 3a summarises the results obtained with particle-signal histograms reminiscent of Gaussian distribution functions. As the particle diameter approaches $\lambda/2 = 260\ nm$ we observe considerable deviation from the Rayleigh scattering approximation, with 250 nm NPs exhibiting less scattering signal than 200 nm NPs, but a simple signal-estimate based on Mie theory nicely reproduces our experimentally observed trend (Figure 3b). Importantly, even though Figure 3a,b are plotted using the cube-root of the measured signal amplitude, to allow for a quantitative representation of the extremely different signals, the camera necessarily measures the scattering intensity which varies by more than five orders of magnitude for the different particles (Figure 3b, inset). To further illustrate the achievable dynamic range we, simultaneously, immobilise the NPs exhibiting the lowest, 20 nm Au, and highest, 200 nm Au, scattering amplitudes on the same sample and record k-scope images as outlined above. Figure 3c shows representative scattering amplitude images that verify that simultaneous detection is indeed possible, albeit a nominal signal difference, as detected by the camera, by approximately $10^6$.

**Dielectric particles and signal-calibration,**

Thus far, we concentrated on measuring Au NPs as their widespread use allows facile cross-platform comparison. The results presented in Figure 3 suggest that quantitative sizing of arbitrary particles should be possible but the challenge is to provide a suitable calibration that allows relating the observed signals to the correct particle diameter. As biological particles are dielectric, we perform calibration experiments using $SiO_2$ particles whose refractive index of $n_{520nm}$=1.461 is somewhere between the extremes of EVs and proteins. Figure 4a shows the histograms obtained for $SiO_2$ particles with diameters of 60 nm, 80 nm and 177 nm in comparison to a measurement of 60 nm Au NPs. The combination of the $SiO_2$ and the Au measurements allows estimating how well the particle size of one species can be determined by using a calibration obtained with a different material. As previously, we use a Mie theory-based simulation to describe the $SiO_2$ signal-dependence (Figure 4b). Simultaneously, we compute the expected signal-curve for Au NPs and scale it by relying on the $SiO_2$ calibration. Figure 4b shows a near-perfect match between the $SiO_2$ and the Au measurements, even though the scaling of the theoretical curves purely relies on the $SiO_2$ data. Given the quality of agreement, albeit the dramatic difference in absolute scattering cross-sections, we suggest that a $SiO_2$ calibration should be well suited for sizing EVs, which exhibit similar cross-sections due to their approximate refractive index of, approximately, n=1.38 [22].

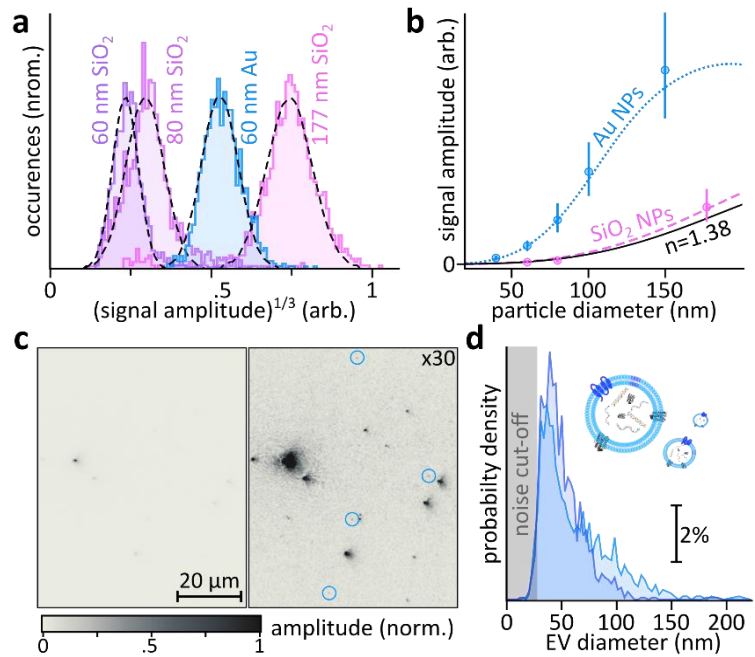

**Figure 4, Absolute size calibration and extracellular vesicle size distribution.** a) SiO$_2$ NP-histograms for three different particle sizes alongside 60 nm Au NPs. b) Theoretical signal dependences normalised based on the SiO$_2$ data (pink) describes the Au NP amplitudes (blue) well. The black line describes the expected signal-scaling for EVs, assuming a refractive index of n=1.38. The error bars correspond to one standard deviation. c) Representative amplitude images of EVs immobilised on mica. d) EV size-distribution histograms for two independent measurements. The steep cut-off towards lower diameters, around 25 nm, is due to signal-to-noise limitations, when only one camera frame is acquired, and not representative of the real particle distribution

**Extracellular vesicles,**

To conclude we perform sizing experiments of ovarian cancer (OVCA) cell-derived EVs. For these initial proof-of-concept demonstrations, we incubate freshly cleaved mica substrates with an aqueous EV suspension for approximately 5 mins and then carefully remove residual water from the surface. As we purely rely on nonspecific binding of the EVs we suspect that we might be selectively detaching some of the larger EVs during the removal process. Despite this non-ideal approach, we clearly observe widely varying signal levels (Figure 4c). The image highlights the extremely heterogeneous nature of EV-samples. The relative scattering amplitudes vary by more than 100-fold, which is equivalent to >10$^4$ in measured signal-intensity. By acquiring multiple images, we obtain EV size-distribution histograms (Figure 4d). Reassuringly, separately prepared samples yield qualitatively similar histograms with some deviations in the larger diameter range as expected based on the considerations given above. Irrespective, our current k-scope is able to detect single EVs with diameters as small as 25 nm, assuming a refractive index of n=1.38 which is on the lower end of the reported values[22–24]. Importantly, the experimentally limiting factor is merely the integration time chosen, which was set to 5 ms in the current experiment, as the mica surface is virtually background-free. Even though we could easily acquire more sensitive data, by averaging multiple images, we opted for the cut-off presented here as we believe that the main limitation in the small size-range is the difficulty to distinguish EVs from other microvesicles and potential contaminations such as protein aggregates or lipoproteins and not the mere detection of scattering particles.

**Summary,**

To summarise, we implemented a k-space holographic microscope (k-scope) that images a sample of interest by interfering a conjugate plane of the microscope's BFP in an off-axis holographic configuration. This Fourier imaging approach dramatically boosts the achievable dynamic range of wide field NP-sensing platforms as the information of all individual particles is projected onto all camera pixels, rather than being localised at a well-defined pixel location. We demonstrated the feasibility and sensitivity of this approach by quantitatively sizing Au NPs with diameters ranging from 20-250 nm. Furthermore, we provided experimental evidence for the k-scope's capability to quantitatively determine the size of an unknown particle-distribution by relying on an absolute size-calibration based on a known set of particles as demonstrated by correctly predicting Au NP-signals based on a $SiO_2$ calibration. Ultimately, we presented initial proof-of-concept work that shows that the k-scope is a promising candidate for a rapid EV size-quantification platform, capable of simultaneously sizing entire EV-populations due to the dramatically improved dynamic range. By coupling our technology with specifically modified sample-substrates and by, furthermore, taking advantage of the enormous miniaturisation-potential of devices based on digital holography we hope to be able to provide a clinically viable sizing-platform in the near future.

**Methods,**

**Sample preparation,** Freshly cleaved mica (V1 grade 0.15-0.21 mm thickness, *Micro to Nano*) is functionalised by adding a 5 µL drop of a 0.1 g/ml solution of PLL-g-PEG (*SuSoS AG*) in PBS and then removing it immediately with a pipette. Following functionalisation, we incubate the sample with 2 µL of a suspension containing citrate-capped Au NPs (*BBI Solutions*), and remove it after a few to tens of seconds, depending on the nominal size of the Au NPs (40-250 nm). The 20 nm Au NP stock-suspension is diluted five-fold prior to incubation. $SiO_2$ NP-samples are fabricated by drying 5 µL of a diluted stock suspension on freshly cleaved mica (60 nm and 80 nm $SiO_2$ NPs were obtained from *nanoComposix*, 177 nm NPs from *micro particles GmBH*).

**Extracellular vesicle preparation from cell culture.** OVCA cells (SkOV3) are cultured in RPMI-1640 medium (*Cellgro*) with 10% fetal bovine serum (FBS) at 37 °C in a humidified atmosphere with 5% $CO_2$. Before EV collection, cells are grown in RPMI with 5% exosome-depleted FBS (Thermo Fisher Scientific) for 48 hours. Supernatants from cell culture media are centrifuged at 300 x g for 5 min to remove cell debris. Supernatant is filtered through a 0.2 µm membrane filter (*Millipore*). Filtered medium is concentrated at 100000 x g for 70 min. After the supernatant has been removed, the EV pellet is washed in PBS and then centrifuged at 100000 x g for 70 min. The EV pellet is resuspended in PBS and stored at -80 °C. EV concentration is determined by Nanoparticle Tracking Analysis (*NanoSight, Malvern*).

**Data analysis,** Prior to performing any data processing, we subtract the camera dark-offset from all images, the following discussion assumes zero dark-offset. We separately acquire an image of the reference wave "reference" by blocking the sample-illumination. We then subtract the references from all acquired holograms and, furthermore divide the resulting image by the square-root of the reference to correct potential amplitude inhomogeneities. Following these corrections, we perform a Fourier transformation and separate the interference information from residual DC components, by cropping the image, as indicated in Figure 2a. We then multiply each image with a spherical lens function to correct for minor de-focus of the BFP on the camera. Finally, we correct the image for defocus and, importantly, potential aberrations, which can be quite severe when transmitting through the freestanding, and often irregular, mica sheets. We select a representative particle within one of the acquired images and isolate it by multiplying the image with a binary circle that eliminates all other image information. The typical circle size is approximately 10 times Nyquist. We inverse Fourier transform the image and remove linear phase ramps, thus moving the particle to DC. We then extract the residual phase information from the complex BFP image and divide the inverse Fourier transformation of all individual images by said phase term. A Fourier transformation of the phase-corrected complex BFP directly yields in-focus, aberration corrected, particle images.

We then identify all particles in all images and fit them using 2D Gaussian functions which yield scattering amplitudes alongside the x/y NP positions. In a final step, we need to correct for the spatially non-uniform illumination profile which directly impacts the scattering amplitude. We reconstruct the beam profile by relying on the amplitudes and positions of all particles. In brief, we generate an image containing all localised particles, normalised to the number of detection events per position and convolve the image with a wide Gaussian function. The area of the resulting beam-estimate is then normalised to unity and all particle-amplitudes are then divided by the beam's amplitude at their respective x/y-position. Particles in the low-amplitude regions of the illumination profile, <10% of the maximum amplitude, are excluded from the analysis as we noted considerable histogram broadening when including said fraction.

**Corresponding Author**

*Matz Liebel, e-mail Matz.Liebel@ICFO.eu. ICFO -Institut de Ciencies Fotoniques, The Barcelona Institute of Science and Technology, 08860 Castelldefels, Barcelona, Spain

**Author Contributions**

The experiments were performed by UOO and ML; UOO processed and analyzed the data; AJ and HL provided and characterized the extracellular vesicles samples. UOO, NFvH and ML wrote the initial manuscript. All authors contributed to the final manuscript and have given approval to the final version of the manuscript.

**Funding Sources**

Authors acknowledge support by the Ministry of Science, Innovation and Universities (MCIU/AEI: RTI2018-099957-J-I00 and PGC2018-096875-B-I00). N.F.v.H. acknowledges the financial support by the European Commission (ERC Advanced Grant 670949-LightNet, ERC Proof-of-Concept Grant 755196-IBIS), Ministry of Science & Innovations ("Severo Ochoa" program for Centers of Excellence in R&D CEX2019-000910-S), the Catalan AGAUR (2017SGR1369), Fundació Privada Cellex, Fundació Privada Mir-Puig, and Generalitat de Catalunya through the CERCA program. H.L. was partly supported by NIH grants (R01CA229777, U01CA233360, R21DA049577) and DoD awards (W81XWH-19-1-0199, W81XWH-19-1-0194).

1. Zheng, G., Horstmeyer, R. & Yang, C. Wide-field, high-resolution Fourier ptychographic microscopy. *Nat. Photonics* **7**, 739–745 (2013).

2. Ozcan, A. & McLeod, E. Lensless Imaging and Sensing. *Annu. Rev. Biomed. Eng.* **18**, 77–102 (2016).

3. Curtis, A. S. G. A Study by Interference Reflection Microscopy. *J. Cell Biol.* **20**, 199–215 (1964).

4. Avci, O., Ünlü, N. L., Özkumur, A. Y. & Ünlü, M. S. Interferometric reflectance imaging sensor (IRIS)—a platform technology for multiplexed diagnostics and digital detection. *Sensors* **15**, 17649–17665 (2015).

5. Huang, Y. F. *et al.* Coherent Brightfield Microscopy Provides the Spatiotemporal Resolution to Study Early Stage Viral Infection in Live Cells. *ACS Nano* **11**, 2575–2585 (2017).

6. Ortega-Arroyo, J. & Kukura, P. Interferometric scattering microscopy (iSCAT): new frontiers in ultrafast and ultrasensitive optical microscopy. *Phys. Chem. Chem. Phys.* **14**, 15625 (2012).

7. Young, G. *et al.* Quantitative mass imaging of single molecules in solution. *Science* **360**, 423–427 (2018).

8. Liebel, M., Hugall, J. T. & van Hulst, N. F. Ultrasensitive label-free nanosensing and high-speed tracking of single proteins. *Nano Lett.* **17**, 1277–1281 (2017).

9. Piliarik, M. & Sandoghdar, V. Direct optical sensing of single unlabelled proteins and super-resolution imaging of their binding sites. *Nat. Commun.* **5**, 4495 (2014).

10. Lin, Y.-H., Wei-Lin, C. & Hsieh, C.-L. Shot-noise limited localization of single 20 nm gold particles with nanometer spatial precision within microseconds. *Opt. Express* **22**, 9159–9170 (2014).

11. van der Pol, E., Boing, A. N., Harrison, P., Sturk, A. & Nieuwland, R. Classification,

Functions, and Clinical Relevance of Extracellular Vesicles. *Pharmacol. Rev.* **64**, 676–705 (2012).

12. Colombo, M., Raposo, G. & Théry, C. Biogenesis, Secretion, and Intercellular Interactions of Exosomes and Other Extracellular Vesicles. *Annu. Rev. Cell Dev. Biol* **30**, 255–89 (2014).

13. Van Niel, G., D'Angelo, G. & Raposo, G. Shedding light on the cell biology of extracellular vesicles. *Nat. Rev. Mol. Cell Biol.* **19**, 213–228 (2018).

14. Mathieu, M., Martin-Jaular, L., Lavieu, G. & Théry, C. Specificities of secretion and uptake of exosomes and other extracellular vesicles for cell-to-cell communication. *Nat. Cell Biol.* **21**, 9–17 (2019).

15. Bachurski, D. *et al.* Extracellular vesicle measurements with nanoparticle tracking analysis–An accuracy and repeatability comparison between NanoSight NS300 and ZetaView. *J. Extracell. Vesicles* **8**, 1596016 (2019).

16. Hillman, T. R., Gutzler, T., Alexandrov, S. A. & Sampson, D. D. High-resolution, wide-field object reconstruction with synthetic aperture Fourier holographic optical microscopy. *Opt. Express* **17**, 7873–7892 (2006).

17. Alexandrov, S. A., Hillman, T. R., Gutzler, T. & Sampson, D. D. Synthetic aperture Fourier holographic optical microscopy. *Phys. Rev. Lett.* **97**, 1–4 (2006).

18. Schnars, U. & Jüptner, W. Direct recording of holograms by a CCD target and numerical reconstruction. *Appl. Opt.* **33**, 179–181 (1994).

19. Cuche, E., Marquet, P. & Depeursinge, C. Spatial filtering for zero-order and twin-image elimination in digital off-axis holography. *Appl. Opt.* **39**, 4070–4075 (2000).

20. Takeda, M., Ina, H. & Kobayashi, S. Fourier-transform method of fringe-pattern analysis for computer-based topography and interferometry. *J. Opt. Soc. Am.* **72**, 156 (1982).

21. Charrière, F. *et al.* Shot-noise influence on the reconstructed phase image signal-to-noise ratio in digital holographic microscopy. *Appl. Opt.* **45**, 7667–7673 (2006).

22. Gardiner, C. *et al.* Measurement of refractive index by nanoparticle tracking analysis reveals heterogeneity in extracellular vesicles. *J. Extracell. Vesicles* **3**, 25361 (2014).

23. de Rond, L. *et al.* Refractive index to evaluate staining specificity of extracellular vesicles by flow cytometry. *J. Extracell. Vesicles* **8**, 1643671 (2019).

24. Rupert, D. L. M. *et al.* Effective Refractive Index and Lipid Content of Extracellular Vesicles Revealed Using Optical Waveguide Scattering and Fluorescence Microscopy. *Langmuir* **34**, 8522–8531 (2018).